\documentclass[pra, showpacs, twocolumn, floatfix]{revtex4}
\usepackage{graphicx}
\usepackage{amsmath, amsfonts, amssymb, bm}

\allowdisplaybreaks[2]

\begin{document}
\title{Quantum entanglement in dense multiqubit systems }

\author{Mihai \surname{Macovei}}
\email{mihai.macovei@mpi-hd.mpg.de}

\author{J\"org \surname{Evers}}
\email{joerg.evers@mpi-hd.mpg.de}

\author{Christoph H. \surname{Keitel}}
\email{keitel@mpi-hd.mpg.de}

\affiliation{Max-Planck-Institut f\"{u}r Kernphysik, Saupfercheckweg
1, D-69117 Heidelberg, Germany}
\date{\today}
\begin{abstract}
The pairwise entanglement of an arbitrary atomic pair randomly 
extracted from a laser-driven dense multiqubit sample in the 
presence of quantum dissipation due to spontaneous emission is 
considered. The dipole-dipole interaction between the particles 
shifts the laser-qubit resonance frequency and consequently 
modifies the quantum entanglement. By means of an appropriate 
tuning of the laser frequency, one can optimize the entanglement 
in this system. For large ensembles, the maximum entanglement 
occurs near the laser parameters where the steady-state sample 
exhibits phase transition phenomena. 
\end{abstract}
\maketitle
\section{Introduction}
The existence of entanglement in multiparticle samples is an  
important fundamental problem \cite{ent_m}. However, this topic is 
difficult due to the fact that a general criterion that quantifies 
the entanglement in such systems does not exist. One may consider 
the state that cannot be represented as a factorized product of 
individual states describing each particle separately as an entangled 
multiparticle state. As a consequence, the condition for entanglement 
of formation of an arbitrary state of two-qubits \cite{cr} and an 
inseparability criterion for continuous variable systems \cite{cv} 
were proposed. Other criteria are, respectively, the entanglement of 
distillation \cite{ed}, the relative entropy of entanglement \cite{re} 
or the so called negativity \cite{ne}. Based on these formulations, 
numerous investigations were performed in order to describe the 
entanglement in various systems. In particular, entanglement induced 
by a single-mode heat environment was shown to occur in \cite{kim} 
while entanglement between two atoms in an overdamped cavity injected 
with squeezed vacuum can be effectively created \cite{gxl}. Two 
initially entangled and afterward not interacting qubits can become 
completely disentangled in a finite time \cite{eb}. However, dark 
periods and revivals of entanglement in a two-qubit system was shown 
to occur via environmental vacuum modes \cite{ft}. An experimental 
demonstration of continuous variable entanglement using cold atoms 
in a high-finesse optical cavity was presented in \cite{cv_ex} while 
experimental evidence of quantum entanglement of a large number of 
photons was proved in Ref.~\cite{ph_ex}. Also, the entanglement of 
formation for an arbitrary two-mode Gaussian state was demonstrated 
in \cite{tm}.

Usually, many-particle systems show critical behaviors such as quantum 
phase transitions \cite{qpt,puri,mek}. The entanglement properties of the 
quantum phase transition in particular systems \cite{qpt1,qpt2} as well 
as their instability \cite{qpt3} have been investigated in detail. 
There, the concurrence \cite{cr}, which characterizes the entanglement 
between two spins (i.e. pairwise entanglement) after tracing out other 
all others, was calculated. This quantity is well-suited for multiparticle 
systems since it does not depend on the two spins selected because all 
spins are completely equivalent \cite{wm}. Moreover, arbitrary symmetric 
multiqubit states are spin squeezed if and only if they are pairwise 
entangled \cite{sp_sq}. Note that spin squeezing was experimentally 
measured in \cite{sp1,sp2}.
\begin{figure}[b]
\includegraphics[width=6.5cm]{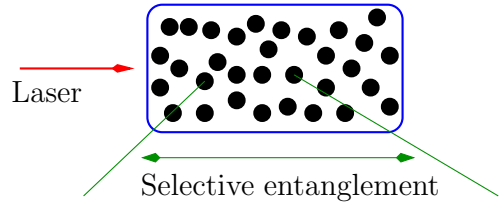}
\caption{\label{fig-1} Schematic of a laser-pumped multiqubit two-level 
ensemble. Our interest is whether or not an arbitrary pair is entangled 
inside the sample.}
\end{figure}

Therefore, in this paper we investigate the pairwise entanglement in a 
laser-pumped dense multiparticle sample which may characterize the quantum 
nature of the system as a whole. The collectivity among the particles is 
mediated through the surrounding electromagnetic field vacuum reservoir 
\cite{nard}. This system is known to exhibit first- or second-order phase 
transitions in the steady-state. We estimate the concurrence for an arbitrary 
atomic pair randomly extracted from the atomic ensemble. The influence of 
the dipole-dipole interaction among particles on the entanglement creation 
is discussed in detail. In particular, the dipole-dipole interactions 
significantly enhance the pairwise entanglement in a two-qubit system. 
For larger samples, the maximum possible value of the concurrence is less 
sensitive to the dipole-dipole interactions. The relationship between the 
entanglement and phase transition phenomena in this system is established. 
We found that the maximum of the pairwise entanglement occurs near the 
critical steady-state behaviors where phase transitions take place for 
larger samples.

The paper is organized as follows. In Section II, we introduce the system 
of interest and solve it in the steady-state. Section III deals with the 
pairwise entanglement in a symmetric multiparticle sample. Section IV 
describes the obtained results. We finalize the article with conclusions 
presented in Section V.

\section{System of interest}
We consider a dense multiqubit two-state ($|e\rangle$ and $|g\rangle$)
sample pumped with a coherent laser field and damped via its interaction 
with the environmental electromagnetic field reservoir (see Fig.~\ref{fig-1}). 
The frequencies of all atoms are identical and equal to $\omega_{0}$, 
while the laser frequency is $\omega_{L}$. The particles are placed in a 
volume with dimensions small compared to the emission wavelength. As a 
consequence, we assume that the dipole-dipole interaction potential, 
$\delta$, is identical for all qubit pairs. Appropriate systems could be 
driven solid-state samples \cite{sss}, superconducting qubits \cite{sq,xmek}, 
pumped multiparticle systems trapped inside optical cavities \cite{ocav} or 
double quantum dot excitonic systems \cite{exs}. Such a model can be 
described in the Born-Markov, the rotating-wave and the electric-dipole 
approximations by the master equation approach \cite{GSA}:
\begin{eqnarray}
\frac{d}{dt}\rho = -i[H_{0},\rho] - \gamma\{[S^{+},S^{-}\rho] 
+ [\rho S^{+},S^{-}]\}, \label{ME}
\end{eqnarray}
where
\begin{eqnarray}
H_{0} = \tilde \Delta S_{z} - \delta S^{+}S^{-} + \Omega(S^{+}+S^{-}),
\label{Hm}
\end{eqnarray}
with $\tilde \Delta=\omega_{0}-\omega_{L} + \delta \equiv \Delta + \delta$.
Here, $\Omega$ is the Rabi frequency and $\gamma$ is the single-particle 
spontaneous emission decay rate. The collective angular momentum operators, 
$S^{\pm}$ and $S_{z}$, are defined in the following way:
\begin{eqnarray}
S^{+} = \sum^{N}_{j=1}\sigma^{+}_{j}, ~~~S^{-} = (S^{+})^{\dagger}, ~~~
S_{z} = \sum^{N}_{j=1}\sigma_{zj}/2, \label{spmz}
\end{eqnarray}
where the raising and lowering operators for each qubit are denoted 
by $\sigma^{+}_{j}=|e\rangle_{j}{}_{j}\langle g|$ and 
$\sigma^{-}_{j}=(\sigma^{+}_{j})^{\dagger}$, while 
$\sigma_{zj}=|e\rangle_{j}{}_{j}\langle e|-|g\rangle_{j}{}_{j}\langle g|$.
The collective operators obey the commutation relations for su(2) algebra, 
i.e. $[S_{z},S^{\pm}]=\pm S^{\pm}$ and $[S^{+},S^{-}]=2S_{z}$.

Since our main interest here is the steady-state entanglement for random 
pairs of qubits extracted from the whole ensemble, we present the 
steady-state solution of the master equation (\ref{ME}) \cite{kl}:
\begin{eqnarray}
\rho_{s} = Z^{-1}\sum^{N}_{n,m=0}C_{nm}(S^{-})^{n}(S^{+})^{m},
\label{sss}
\end{eqnarray}
where $C_{nm}=(-1)^{n+m}\alpha^{-n}(\alpha^{*})^{-m}a_{nm}$ with 
\begin{eqnarray}
\label{anm}
a_{nm}=\frac{\Gamma(1+n+\beta)\Gamma(1+m+\beta^{*})}{n!m!\Gamma(1+\beta)\Gamma(1+\beta^{*})}.
\end{eqnarray}
Further, $\alpha=i\Omega/(\gamma+i\delta)$ and $\beta=i\tilde\Delta/(\gamma+i\delta)$ 
while $Z$ is chosen in such a way that ${\rm Tr\{\rho_{s}\}}=1$. The trace can be 
performed using the relations
\begin{eqnarray}
S^{+}|s,l\rangle=\sqrt{(s-l)(s+l+1)}|s,l+1\rangle, \nonumber \\
S^{-}|s,l\rangle=\sqrt{(s+l)(s-l+1)}|s,l-1\rangle,
\label{sl}
\end{eqnarray}
where the collective Dicke states $|s,l\rangle$, with $s=N/2$ and 
$-s \le l \le s$, are the eigenstates for the operator $S_{z}$ and 
the operator of the total "spin" $S^{2}$:
\begin{eqnarray}
S_{z}|s,l\rangle=l|s,l\rangle, \nonumber \\
S^{2}|s,l\rangle=s(s+1)|s,l\rangle.
\end{eqnarray}
Therefore, 
\begin{eqnarray}
Z=\sum^{N}_{n=0}a_{nn}|\alpha|^{-2n}\frac{(N+n+1)!(n!)^{2}}{(N-n)!(2n+1)!}.
\label{Z}
\end{eqnarray}
The steady-state atomic variables of interested can be obtained 
using Eqs.~(\ref{sss}-\ref{Z}):
\begin{eqnarray}
&{}&\langle(S^{+})^{p}(S_{z})^{r}(S^{-})^{f} \rangle = Z^{-1}\sum^{N}_{n=max\{f,p\}}
C_{n-f,n-p} \nonumber \\
&\times&\sum^{N-n}_{m=0}\frac{(N-m)!(m+n)!}{(N-m-n)!m!}(N/2-m-n)^{r},
\label{ssv}
\end{eqnarray}
where $\{p,r,f\} \in \{0,1,2, \cdots, N\}$. The exact steady-state results 
for large samples show that, for $\Delta=\delta=0$, the steady state 
averages are a continuous function of the driving field parameter 
$2\Omega/(N\gamma)$ but their derivatives with respect to that parameter 
are discontinuous at $2\Omega/(N\gamma)=1$. This behavior is reminiscent 
of a second-order phase transition. If $\Delta \not = 0$ but $\delta=0$ 
there is not such a critical phenomenon \cite{puri}. However, when 
$\{\Delta,\delta\}\not=0$ the steady state solution (\ref{sss}) 
predicts a first-order phase transition for large systems \cite{kl,puri}.

Further, we shall use the expectation values of the collective operators 
given by Eq.~(\ref{ssv}) to estimate the pairwise entanglement among any 
pairs randomly extracted from the multiqubit ensemble. The relationship 
between the entanglement and above mentioned phase transition phenomena 
will be established.

\section{Pairwise entanglement}
For a mixed state of qubits $\{a,b\} \in \{1, \cdots,N\}$ with density 
matrix $\rho_{ab}$, the concurrence $C$ is defined as 
\begin{eqnarray}
C={\rm max}\{0,\lambda_{1}-\lambda_{2}-\lambda_{3}-\lambda_{4}\},
\label{C}
\end{eqnarray}
where the quantities $\lambda_{i}$~($i \in \{1,2,3,4\}$) are the square 
roots of the eigenvalues of the matrix product
\begin{eqnarray}
R=\rho_{ab}(\sigma_{ay}\otimes\sigma_{by})\rho^{*}_{ab}(\sigma_{ay}\otimes\sigma_{by}),
\label{R}
\end{eqnarray}
in descending order. Here, $\rho^{*}_{ab}$ denotes complex conjugation 
of $\rho_{ab}$, and $\sigma_{iy}$ are Pauli matrices for the two-level 
systems $\bigl(i \in \{a,b\}\bigr)$. The values of the concurrence range 
from zero for an unentangled state to unity for a maximally entangled 
two-particle state \cite{cr}. Note that we do not rescale the obtained 
concurrence with the number of atoms $N$, in contrast to some previous 
works \cite{qpt2,wm}.
\begin{figure}[t]
\includegraphics[width=7cm]{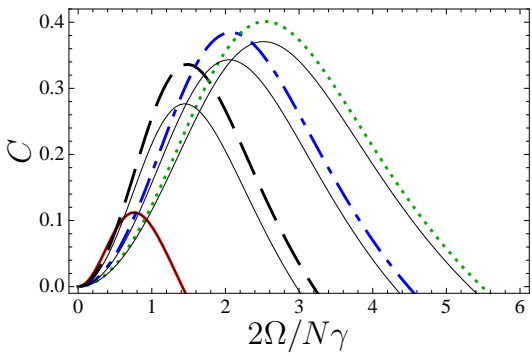}
\caption{\label{fig-2}(color online) The concurrence $C$ as function of 
$2\Omega/(N\gamma)$ for $N=2$ and $\tilde \Delta+\delta=0$. 
The solid curve is for $\delta/\gamma=0$, the long-dashed line 
corresponds to $\delta/\gamma=5$, the dashed-dotted line is for 
$\delta/\gamma=10$ while the dotted curve is for $\delta/\gamma=15$. Thin
solid lines show corresponding results for $C_{ref}^{(1)}$ defined in 
Eq.~(\ref{cfit}).}
\end{figure}

The state of two qubits which is symmetric under the exchange of the sub-systems
can be represented in the basis $\{|e_{a}e_{b}\rangle,|e_{a}g_{b}\rangle,
|g_{a}e_{b}\rangle,|g_{a}g_{b}\rangle \}$ as follows:
\begin{eqnarray}
\rho_{ab} = \left( \begin{array}{cccc}
\rho_{11} & \rho_{12} & \rho_{13} & \rho_{14}\\
\rho_{21} & \rho_{22} & \rho_{23} & \rho_{24}\\
\rho_{31} & \rho_{32} & \rho_{33} & \rho_{34} \\ 
\rho_{41} & \rho_{42} & \rho_{43} & \rho_{44}
\end{array} \right). 
\label{rab}
\end{eqnarray}
For symmetric multiparticle systems like ours, the density matrix $\rho_{ab}$ 
and hence the concurrence do not depend on the specific choice of 
qubits $a$ and $b$. Therefore, the matrix elements of $\rho_{ab}$ can 
be determined by:
\begin{align}
\rho_{11}&=\frac{N^{2}-2N+4\langle S^{2}_{z}\rangle +4(N-1)\langle S_{z}\rangle}
{4N(N-1)}, \nonumber \\
\rho_{12}&=\rho_{13}=\frac{N\langle S^{+}\rangle + 2\langle S^{+}S_{z}\rangle}
{2N(N-1)}, \nonumber \\
\rho_{14}&=\frac{\langle (S^{+})^{2}\rangle}{N(N-1)}, ~~~\rho_{21}=(\rho_{12})^{*}, \nonumber \\
\rho_{22}&=\rho_{23}=\frac{N^{2}-4\langle S^{2}_{z}\rangle}{4N(N-1)}, \nonumber \\
\rho_{24}&=\frac{\langle S^{+}\rangle(N-2)-2\langle S^{+}S_{z}\rangle}{2N(N-1)}, \nonumber \\
\rho_{31}&=(\rho_{13})^{*}, ~~~ \rho_{32}=(\rho_{23})^{*}, ~~~\rho_{33}=\rho_{22}, \nonumber \\
\rho_{34}&=\rho_{24}, ~~~ \rho_{41}=(\rho_{14})^{*}, ~~~\rho_{42}=(\rho_{24})^{*}, \nonumber \\
\rho_{43}&=(\rho_{34})^{*}, \nonumber \\
\rho_{44}&=\frac{N^{2}-2N + 4\langle S^{2}_{z}\rangle - 4(N-1)\langle S_{z}\rangle}
{4N(N-1)}.
\label{em}
\end{align}
As it can be observed here, the two-particle density matrix is expressed 
in terms of expectation values of collective operators. The approach 
is valid when one has a symmetric state of $N$ two-level particles, 
where by symmetric, we mean symmetry under any permutation of the 
qubits \cite{wm,sp_sq}. Thus, the relations (\ref{ssv}-\ref{em}) will 
help us to estimate the pairwise entanglement of an arbitrary qubit 
pair inside the multiparticle ensemble. 

We also note that it was found in the literature that for density matrices 
which due to the studied model are restricted to the simpler case
$\rho_{12}=\rho_{13}=\rho_{24}=\rho_{34}=0$, a compact analytic expressions for
the concurrence can be found, which is given by~\cite{zubairy}
\begin{align}
C_{ref} = \max \left \{0, C_{ref}^{(1)}, C_{ref}^{(2)} \right\}\,,\\
C_{ref}^{(1)} = 2 (|\rho_{14}| - \sqrt{\rho_{22}\rho_{33}})\,,\\
C_{ref}^{(2)} = 2 (|\rho_{23}| - \sqrt{\rho_{11}\rho_{44}})\,.
\end{align}
%
We will see that $C_{ref}^{(1)}$ which for the density matrix elements
in Eq.~(\ref{em}) evaluates to
\begin{align}
\label{cfit}
C_{ref}^{(1)} = 2 (|\rho_{14}| - \rho_{22})
\end{align}
will turn out to fit most of our results well. In the next Section, 
we turn to a  characterization of the two-qubit concurrence in various parameter 
ranges.
\begin{figure}[t]
\includegraphics[width=7cm]{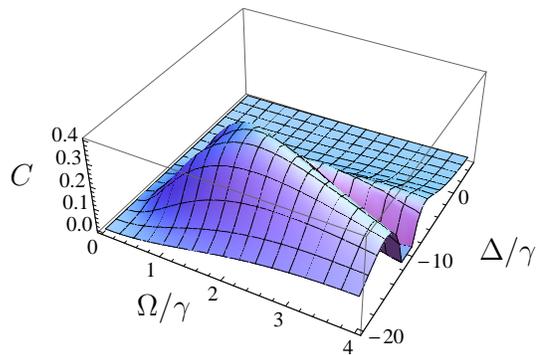}
\caption{\label{fig-3}(color online) The concurrence $C$ as function of 
$\Omega/\gamma$ and $\Delta/\gamma$. Here $N=2$ and $\delta/\gamma=5$.}
\end{figure}

\section{Results and Discussion}
In Figure \ref{fig-2}, we plot the steady-state concurrence for a driven 
two-qubit system. The solid curve describes the case when dipole-dipole 
interaction among the particles is absent while the laser field is on 
resonance with the qubit frequency. The other curves characterize the 
situations where the dipole-dipole interaction increases while the laser 
field frequency obeys the condition $\tilde \Delta + \delta=0$. These 
behaviors can be understood in the following way. In the limit of small 
interparticle distance, the two-particle system is equivalent to a 
three-state ladder system: $|g_{a}g_{b}\rangle$ being the ground state, 
$|s\rangle=(|e_{a}g_{b}\rangle + |e_{b}g_{a}\rangle)/\sqrt{2}$ 
is the intermediate level and the excited state is $|e_{a}e_{b}\rangle$.
The intermediate level $|s\rangle$ is shifted by the dipole-dipole 
interaction potential from the qubit frequency $\omega_{0}$. Therefore, 
in order to be on resonance $\tilde \Delta + \delta=0$, one has to 
modify the laser frequency as shown in Fig.~\ref{fig-2} by the 
interrupted curves. As a result, the maximum of the concurrence moves to 
higher laser strengths. To further characterize the two-qubit system, 
in Fig.~\ref{fig-3}, we fix the dipole-dipole interaction potential, 
i.e. $\delta/\gamma=5$, and change the frequency of the laser field.
Interestingly, the concurrence can be further increased in comparison 
to the one shown in Fig.~\ref{fig-2} by the long-dashed curve, where
$C_{max} \approx 0.34$ is achieved. For example, in Fig.~\ref{fig-3} the maximum 
concurrence is $C_{max} \approx 0.4$ when $\Delta/\gamma=-12$ and 
$\Omega/\gamma \approx 1.8$. Moreover, the concurrence shows oscillatory 
behaviors for larger field intensities as a function of $\Delta$ (see Fig.~\ref{fig-3}). 
\begin{figure}[t]
\includegraphics[width=7cm]{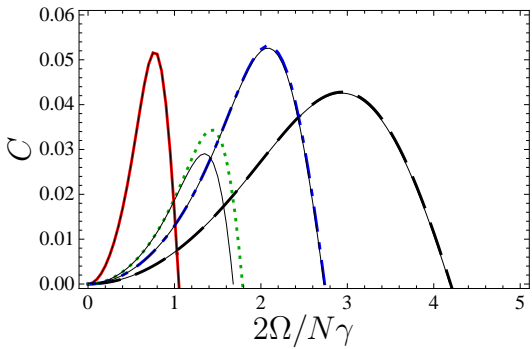}
\caption{\label{fig-4}(color online) The pairwise entanglement $C$ as 
function of $2\Omega/(N\gamma)$ for $N=6$. The solid red line is for 
$\delta=\Delta=0$, the long-dashed black line stands for 
$2\delta/(N\gamma)=1$ and $2\Delta/(N\gamma)=-0.25$, the dashed-dotted 
blue curve for $2\delta/(N\gamma)=1$ and $2\Delta/(N\gamma)=-1.4$, 
while the dotted green curve corresponds to $2\delta/(N\gamma)=1$ and 
$2\Delta/(N\gamma)=-2$. 
Thin solid lines show corresponding results for $C_{ref}^{(1)}$ defined in 
Eq.~(\ref{cfit}).}
\end{figure}

\begin{figure}[t]
\includegraphics[width=7cm]{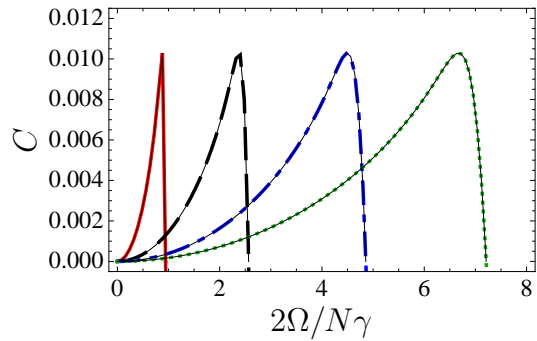}
\caption{\label{fig-5}(color online) The concurrence $C$ as function of 
$2\Omega/(N\gamma)$ for $N=50$. The solid red line is for $\delta=\Delta=0$, 
the long-dashed black line stands for $2\delta/(N\gamma)=0.1$ and 
$2\Delta/(N\gamma)=-0.1$, the dashed-dotted blue curve for $2\delta/(N\gamma)=0.2$ 
and $2\Delta/(N\gamma)=-0.2$, while the dotted green curve corresponds to 
$2\delta/(N\gamma)=0.3$ and $2\Delta/(N\gamma)=-0.3$. 
Thin solid lines show corresponding results for $C_{ref}^{(1)}$ defined in 
Eq.~(\ref{cfit}).}
\end{figure}

\begin{figure}[t]
\includegraphics[width=7cm]{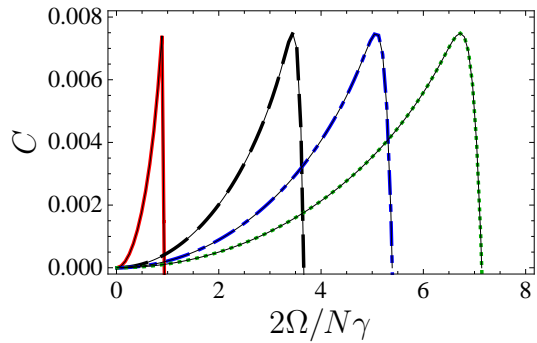}
\caption{\label{fig-6}(color online) The concurrence $C$ as function of 
$2\Omega/(N\gamma)$ for $N=74$. The solid red line is for $\delta=\Delta=0$, 
the long-dashed black line stands for $2\delta/(N\gamma)=0.1$ and 
$2\Delta/(N\gamma)=-0.1$, the dashed-dotted blue curve for 
$2\delta/(N\gamma)=0.15$ and $2\Delta/(N\gamma)=-0.15$, while the dotted 
green curve corresponds to $2\delta/(N\gamma)=0.2$ and $2\Delta/(N\gamma)=-0.2$. 
Thin solid lines show corresponding results for $C_{ref}^{(1)}$ defined in 
Eq.~(\ref{cfit}).}
\end{figure}

Further, we focus on the pairwise entanglement of a pair of qubits 
arbitrarily extracted from a multiparticle sample. We start with a 
moderate ensemble containing several particles. Figure (\ref{fig-4}) 
depicts the pairwise entanglement for such a system with $N=6$ qubits.
There, the solid line stands for $\delta=\Delta=0$ while all others 
are for a fixed dipole-dipole interaction and various laser field 
detunings. Similar to the two-qubit case, the concurrence can be 
enhanced for moderate systems by modifying the external controllable 
laser parameters, i.e. $\{\Omega,\Delta\}$, as well as the dipole-dipole 
interactions $\delta$. For a fixed dipole-dipole interaction, the 
concurrence increases while changing the qubit-laser detuning and then 
decreases with an additional increase of the detuning (see Fig.~\ref{fig-4}). 
For larger samples, i.e. $N \gg 1$, the maximum possible value of the 
concurrence is less sensitive to the dipole-dipole interparticle 
interactions, i.e. has almost the same value as for the case of 
$\delta=\Delta=0$ (see Fig.~\ref{fig-5} and Fig.~\ref{fig-6}). Here, 
an important issue is that for $\delta=\Delta=0$ and larger samples 
the maximum of the concurrence occurs near $2\Omega/(N\gamma) =1$ \cite{qpt1} 
and drops there abruptly (see Fig.~{\ref{fig-5}} and Fig.~{\ref{fig-6}}). 
At this critical point, i.e. $2\Omega/(N\gamma) =1$ and $N \gg 1$, there exists a 
second-order phase transition behavior of the collective atomic 
variables \cite{puri}. In the presence of the dipole-dipole interactions, 
i.e. $\delta \not =0$, the maximum of the pairwise entanglement occurs at 
particular laser frequencies and may shift towards larger values of 
$2\Omega/(N\gamma)$ depending on the laser-qubit detuning. Here again, the 
maximum of concurrence suddenly goes to zero at some critical values of 
$2\Omega/(N\gamma)$ (see Fig.~{\ref{fig-5}} and Fig.~{\ref{fig-6}}). However, 
at these critical values of the pumping parameter, a first-order phase 
transition of the collective atomic variables takes place \cite{kl}. 
Therefore, one can conjecture that always for larger samples and in the 
presence of quantum dissipation due to spontaneous emission, the maximum 
of pairwise entanglement occurs nearby the critical behaviors of the 
collective atomic variables, i.e. phase transitions.

It is interesting to see that in particular for larger samples $(N>1)$ and
for parameters close to those which maximize the obtained concurrence,
the reference $C_{ref}^{(1)}$ in Eq.~(\ref{cfit}) fits our results very well.
In Figs.~\ref{fig-5} and \ref{fig-6}, no difference to the exact results can be seen,
while the coincidence is equally good in Fig.~\ref{fig-4} except for the dotted 
curve with detuning away from the value leading to the optimum concurrence.
For the two-atom case, however, $C_{ref}^{(1)}$ only reflects our numerical
results for the case of absent dipole-dipole interaction, see Fig.~\ref{fig-2}.
Note that $C_{ref}^{(2)}$ does not indicate entanglement in any of
our examples. The good agreement with $C_{ref}^{(1)}$ suggests that it is in fact
entanglement between the collective ground and excited states $|4\rangle$
and $|1\rangle$ which leads to the observed non-zero concurrence, rather
than entanglement between the two intermediate states $|2\rangle$ and $|3\rangle$.
This is somewhat remarkable, as the entangled symmetric state 
$(|2\rangle+|3\rangle)/\sqrt{2}$ naturally arises as an eigenstate of the 
coherent dynamics of dipole-dipole interacting two-level systems.

\section{Summary}
In summary, we investigated the pairwise entanglement of a randomly 
extracted qubit-pair from a laser-driven dense multiparticle ensemble. 
The whole system is damped collectively via its interaction with the 
environmental electromagnetic field vacuum modes. We started with a 
two-qubit system where the dipole-dipole interaction significantly 
enhances the concurrence. Larger values for the concurrence in such 
systems can be obtained by adjusting the laser field detuning 
$\tilde \Delta$ around the frequency shift due to dipole-dipole interaction. 
Intense driving laser fields lead to an oscillatory behavior of the concurrence in 
an off resonant two-particle sample as a function of the detuning. 
For moderate sample sizes, the concurrence 
can be further increased by suitably adjusting the laser frequency and dipole-dipole 
interactions, similar to the two-qubit systems. Interestingly, the pairwise entanglement 
goes abruptly to zero for larger samples, at some particular values of the 
pumping parameter $2\Omega/(N\gamma)$. At these critical values, phase 
transition takes place.


\end{document}